\documentclass{article}
\usepackage{spconf,amsmath,graphicx,amssymb,amsthm,bbm}

\usepackage{xcolor}
\usepackage{svg, import, subfiles, tikz, calc, ifthen, url, hyperref}
\usetikzlibrary{math}
\graphicspath{{Figures/}{../Figures/}}

\newcommand{\ind}[1]{^{\left(#1\right)}}
\newcommand{\vect}[1]{\mathbf{#1}}
\newcommand{\mat}[1]{\mathbf{#1}}
\newcommand{\spikes}[1]{\left\lbrace t_\ell\ind{#1}, \ell = 1\cdots n^{(#1)}_{\mathrm{spikes}} \right\rbrace}
\newcommand{\spikesSingle}{\left\lbrace t_\ell, \ell \in \mathbb{Z} \right\rbrace}

\newcommand{\proj}[1]{\mathcal{P}_{\vect{#1}}}
\newcommand{\set}[1]{\mathcal{C}_{\vect{#1}}}
\newcommand{\setom}{\mathcal{C}_{\Omega,L}}
\newcommand{\projom}{\mathcal{P}_{\Omega,L}}

\title{Sampling and Reconstruction of Mixed Bandlimited Signals using a Set of Spiking Integrate and Fire Neurons}
\title{Encoding and Decoding Mixed Bandlimited Signals using Spiking Integrate-and-Fire Neurons}

\def\FUND{This work was supported by the Swiss National Science Foundation grant number 200021\_181978/1, ``SESAM - Sensing and Sampling: Theory and Algorithms''.}
\name{Karen Adam, Adam Scholefield and Martin Vetterli\thanks{\FUND}}
\address{School of Computer and Communication Sciences\\
    Ecole Polytechnique F\'{e}d\'{e}rale de Lausanne (EPFL)}

\newtheorem{theorem}{Theorem}[section]
\newtheorem{lemma}{Lemma}[section]

\newtheorem{definition}{Definition}[section]

\begin{document}
%\ninept
%
\maketitle
\begin{abstract}
Conventional sampling focuses on encoding and decoding bandlimited signals by recording signal amplitudes at known time points. Alternately, sampling can be approached using biologically-inspired schemes. Among these are integrate-and-fire time encoding machines (IF-TEMs). They behave like simplified versions of spiking neurons and encode their input using spike \textit{times} rather than amplitudes. %Moreover, 
When multiple of these neurons jointly process a set of mixed signals, they form one layer in a feedforward spiking neural network. In this paper,  we investigate the encoding and decoding potential of such a layer. 
We propose a setup to sample a set of bandlimited signals formed by summing a finite number of sincs, by mixing them and sampling the result  using different IF-TEMs. We provide conditions for perfect recovery of the set of signals from the samples in the noiseless case, and suggest an algorithm to perform the reconstruction.
\end{abstract}


\section{Introduction}
While most sampling schemes encode amplitude as a function of time, time encoding, as the name suggests, encodes input signals using signal-dependent time points. In this sense, time encoding machines can be compared to neurons which encode their inputs in spikes, the \emph{timings} of which contains the information about the input~\cite{burkitt2006review1, gerwinn2011reconstructing}.

Time encoding and decoding has been studied for different signal classes, from bandlimited signals~\cite{lazar2004perfect, feichtinger2012approximate, florescu2015novel} to the more general shift-invariant subspaces~\cite{gontier2014sampling} and even to some classes of finite rate of innovation signals~\cite{alexandru2019reconstructing}, mostly by relating the recorded times to irregular samples and performing reconstruction from irregular samples~\cite{feichtinger1994theory, aldroubi2002non}. 

Furthermore, time encoding can take different forms. The most general one is described by Gontier~\cite{gontier2014sampling}, consisting of comparing a filtered version of the input signal to a test function and recording the time points where the two match.

More specific definitions of time encoding can adopt a more biological approach. Time encoding machines can resemble integrate-and-fire neurons with perfect integrators~\cite{adam2019sampling}, or leaky integrate-and-fire neurons with refractory periods~\cite{lazar2005multichannel} or even Hodgkin-Huxley neurons for more biological resemblence~\cite{lazar2010population}.
Moreover, time encoding machines can be used in different configurations, such as single-signal single-channel encoding, or single-signal multi-channel encoding which improves signal reconstruction~\cite{alexandru2019reconstructing, adam2019sampling,lazar2005multichannel}.

\begin{figure*}[tb]
		\centering
        \begin{center}
\begin{tikzpicture}[scale = 0.85]
% TEMs
% TEMs 1, 2, 3
\foreach \n in {0,1,2,3}
{
    
    \tikzmath{
        int \indn;
        \indn= \n +1;
        \startx = \TEMStartCoordX;
        \starty = \TEMStartCoordY - \n*(\TEMRectHeight+\TEMVertSpacing);
    }
    \ifnum \n = 3
        \tikzmath{
            \starty = \TEMStartCoordY - 4*(\TEMRectHeight+\TEMVertSpacing);
        }
    \fi
    \tikzmath{
        \endx = \TEMStartCoordX + \TEMRectWidth;
        \endy = \starty + \TEMRectHeight;
        \linestartX = (\startx - \yToTEMLineLength);
        \lineY = (\starty+\endy)/2);
        \TEMLabelX = (\startx+\endx)/2;
    }
    \draw (\startx, \starty) rectangle (\endx, \endy);
    \ifnum \n < 3
        \draw (\TEMLabelX, \lineY) node {TEM$\ind{\indn}$};
    \else
        \draw (\TEMLabelX, \lineY) node {TEM$\ind{I}$};
    \fi
    \draw (\linestartX, \lineY) -- (\startx, \lineY) ;
    \draw (\startx+\TEMRectWidth, \lineY) -- (\startx+\TEMRectWidth+\TEMOutputLineSpan, \lineY);
    \tikzmath{
        \startSpikex = \startx+\TEMRectWidth+\TEMOutputLineSpan + \TEMOutSpikeHorSpace;
        \startSpikey = \lineY - \TEMOutSpikeVertSpace;
    }
    \ifnum \n = 0
        \foreach \loc in {0.1,0.4,0.6,1,1.5, 1.7}
        {
            \draw (\startSpikex + \loc, \startSpikey) -- (\startSpikex + \loc, \startSpikey + \SpikeHeight);
        }
    \fi
    \ifnum \n = 1
        \foreach \loc in {0.2,0.7,1.1, 1.6}
        {
            \draw (\startSpikex + \loc, \startSpikey) -- (\startSpikex + \loc, \startSpikey + \SpikeHeight);
        }
    \fi
    \ifnum \n = 2
        \foreach \loc in {0.15,0.5,0.8,1,1.2,1.7}
        {
            \draw (\startSpikex + \loc, \startSpikey) -- (\startSpikex + \loc, \startSpikey + \SpikeHeight);
        }
    \fi
    \ifnum \n = 3
        \foreach \loc in {0.3,1, 1.6}
        {
            \draw (\startSpikex + \loc, \startSpikey) -- (\startSpikex + \loc, \startSpikey + \SpikeHeight);
        }
    \fi
    
    \ifnum \n <3
    \draw (\linestartX, \lineY) node[anchor = east] {$y\ind{\indn}(t)$};
    \draw (\startSpikex, \startSpikey) -- (\startSpikex + \TEMSpikeLineSpan, \startSpikey) node[anchor = south west] {$\quad \spikes{\indn}$};
    \else
    \draw (\linestartX, \lineY) node[anchor = east] {$y\ind{I}(t)$};
    \draw (\startSpikex, \startSpikey) -- (\startSpikex + \TEMSpikeLineSpan, \startSpikey) node[anchor = south west] {$\quad \spikes{I}$};
    \fi
}
    
% dotdotdots
\foreach \n in {-1,0,1}
    \tikzmath{
        \centerx = \TEMStartCoordX + \TEMRectWidth/2;
        \centery = \TEMStartCoordY - 3*(\TEMRectHeight+\TEMVertSpacing) + 0.5*\TEMRectHeight +\n*(\dotdotdotspace+\dotdotdotsize);
    }
    \fill (\centerx, \centery) circle (\dotdotdotsize);
    
% TEM M
% \tikzmath{
%     \startx = \TEMStartCoordX;
%     \starty = \TEMStartCoordY - 4*(\TEMRectHeight+\TEMVertSpacing);
%     \endx = \TEMStartCoordX + \TEMRectWidth;
%     \endy = \starty + \TEMRectHeight;
%     \linestartX = (\startx - \yToTEMLineLength);
%     \lineY = (\starty+\endy)/2);
% }
% \draw (\startx,\starty) rectangle (\endx, \endy);
% \draw (\linestartX, \lineY) -- (\startx, \lineY);

\tikzmath{
    \MTEMTotalHeight = 4*\TEMRectHeight + 3*\TEMVertSpacing;
}

% Draw all connections to x, and from x to y
\foreach \n in {0,1,2}
{
    \tikzmath{
        \lineStartX = \TEMStartCoordX - \yToTEMLineLength - \XtoYLineXSpan - \YHorSpace;
    }
        \ifnum \n < 2
        \tikzmath{\lineStartY = \TEMStartCoordY - \n*(\TEMRectHeight+\TEMVertSpacing) - \TEMVertSpacing/2;}
        \else
        \tikzmath{\lineStartY = \TEMStartCoordY - (\n+1)*(\TEMRectHeight+\TEMVertSpacing) - \TEMVertSpacing/2;}
        \fi

    \foreach \m in {0,1,2,3}
    {
        \tikzmath{\lineEndX = \TEMStartCoordX - \yToTEMLineLength -\YHorSpace;}
        \ifnum \m < 3
        \tikzmath{\lineEndY = \TEMStartCoordY - \m*(\TEMRectHeight+\TEMVertSpacing) +\TEMRectHeight/2;}
        \else
        \tikzmath{\lineEndY = \TEMStartCoordY - (\m+1)*(\TEMRectHeight+\TEMVertSpacing) +\TEMRectHeight/2;}
        \fi
        \tikzmath{
        \aX = 2*\lineStartX/4 +\lineEndX/2 ;
        \aY = 2*\lineStartY/4 +\lineEndY/2;
        }
        
        \tikzmath{
            int \indn;
            \indn= \n +1;
            int \indm;
            \indm= \m +1;
        }
        \ifnum \n =0
            \ifnum \m <3
                \draw[highlightgreen, thick]  (\lineEndX, \lineEndY) -- (\lineStartX, \lineStartY) node[midway, above, font=\boldmath, sloped] {$a_{\indm,\indn}$};
            \else
                \ifnum \m = 3
                \draw[highlightgreen, thick]  (\lineEndX, \lineEndY) -- (\lineStartX, \lineStartY) node[midway, above, font=\boldmath, sloped] {$a_{I,\indn}$};
                \fi
            \fi
            
        \else
            \draw (\lineStartX, \lineStartY) -- (\lineEndX, \lineEndY);
        \fi
    }
    \tikzmath{
        int \indn;
        \indn= \n +1;
    }
    \ifnum \n < 2
    \draw (\lineStartX-\XHorSpace-\XInLineSpan, \lineStartY) -- (\lineStartX-\XHorSpace, \lineStartY) node[anchor=west] {$x\ind{\indn}(t)$};
    \else
    \draw (\lineStartX-\XHorSpace-\XInLineSpan, \lineStartY) -- (\lineStartX-\XHorSpace, \lineStartY) node[anchor=west] {$x\ind{J}(t)$};
    \fi
}

\end{tikzpicture}
\end{center}
    \vspace{-1.5em}
	\caption{Sampling setup: $J$ input signals $x\ind{j}(t)$, $j=1\cdots J$ are mixed using a matrix $\mat{A}$ and produce signals $y\ind{i}(t)$, $i=1\cdots I$. Each $y\ind{i}(t)$ is then sampled using a time encoding machine TEM$\ind{i}$ which produces spike times $\spikes{i}$.}
	\label{fig: Figure 1}
	\end{figure*}
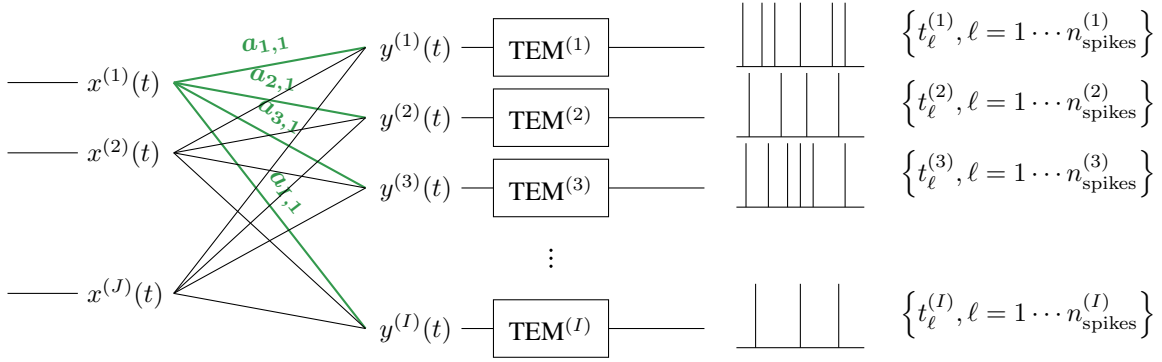

In the present paper, we consider time encoding of multiple bandlimited signals using multiple time encoding machines. We assume that the signals can be written as a finite sum of sincs and that they are mixed before being input to machines with different spiking rates, as depicted in Fig.~\ref{fig: Figure 1}.

Our goal here is to understand how information is encoded in such a network, when it can be fully recovered, and how to perform the recovery.
If each time encoding machine acts like an integrate-and-fire neuron, our setup resembles a single feedforward layer in a spiking neural network. 

We will see how the total number of spikes of this layer should relate to the number of degrees of freedom of the input signals to ensure perfect recovery. Furthermore, time encoding machines or neurons that spike too little can be compensated for by others that spike more frequently, but only up to a certain extent.

First, we present the sampling setup for mixed multi-channel time encoding. We then give a bound for reconstructability of the input signals that is dependent on the number of degrees of freedom of these signals. Finally, we present a recursive reconstruction algorithm and provide some simulation results.
	
\section{Previous work}

Sampling and reconstruction of single bandlimited signals using one or more time encoding machines (TEMs) has been studied. Initial results for single-signal single-channel encoding were established by Lazar and T\'oth~\cite{lazar2004perfect}.

They assume that the input is a signal $x(t)$, which is $2\Omega$-bandlimited in  $L^2(\mathbb{R})$ and bounded such that $|x(t)|\leq c$ for some $c\in\mathbb{R}$, and the TEM has parameters $\kappa$, $\delta$ and $b$, with $b>c$.
\begin{definition}
A signal $x(t)$ is $2\Omega$-bandlimited if its Fourier transform $F_x(\omega)$ satisfies $F_x(\omega) = 0, \forall \, |\omega|>\Omega$.
\end{definition}

\begin{definition}
    An \textit{integrate-and-fire time encoding machine} (IF-TEM) with parameters $\kappa$, $\delta$, and $b$ takes an input signal $x(t)$, adds $b$ to it and integrates the result, scaled by $1/\kappa$, until a threshold $\delta$ is reached. Once this threshold is reached, a time is recorded, the value of the integrator resets to $-\delta$ and the mechanism restarts. We say that the machine spikes at the integrator reset and call the recorded time $t_k$ a \textit{spike time}. 
\end{definition}
The circuit of an IF-TEM is depicted in Fig.~\ref{fig:TEM circuit}.

Lazar and T\'oth showed that if such an $x(t)$ is sampled noiselessly using an IF-TEM and if $\Omega< \pi\left(b-c\right)/\left(2\kappa\delta\right)$, then $x(t)$ can be perfectly recovered from samples $\spikesSingle$ using a recursive algorithm~\cite{lazar2004perfect, lazar2004timerefractory}.

We extended the work to understand single-signal $I$-channel time encoding by building on the approach of Lazar and T\'oth~\cite{lazar2004perfect}.
We showed that if a $2\Omega$-bandlimited, $c$-bounded signal $x(t)$ is sampled noiselessly using $I$ IF-TEMs with the same parameters $\kappa$, $\delta$ and $b$ but with nonzero shifts between their integrators, then $x(t)$ can be reconstructed from its samples $t\ind{i}_\ell$ using a recursive algorithm if $
    \Omega< I\pi\left(b-c\right)/\left(2\kappa\delta\right)$~\cite{adam2019sampling}. Essentially, if a $2\Omega$-bandlimited signal can be reconstructed using one TEM, then a $2I\Omega$-bandlimited signal can be reconstructed using $I$ TEMs with the same parameters.
    
In this paper, we further extend the setup to allow for $J$-signal $I$-channel time encoding, where $I, J \in \mathbb{N^+}$.

\section{Sampling Setup and Reconstructability Constraints}

\subsection{Sampling Setup}
We now give a brief overview of the sampling setup as depicted in Fig.~\ref{fig: Figure 1}. Further details are provided in Section~\ref{sec: Setup constraints}.

Our setup assumes that we are interested in encoding $J$ $2\Omega$-bandlimited signals $x\ind{1}(t)$,  $x\ind{2}(t)$, $\cdots$, $x\ind{J}(t)$ where
\begin{equation} \label{eq: condition on x's}
    x\ind{j}(t) = \sum_{k=1}^{K} c_{jk} \, \mathrm{sinc}_\Omega (t-t_k), \quad \forall j=1\cdots J,
\end{equation}
with $\mathrm{sinc}_\Omega(t) = \sin(\Omega t)/(\pi t)$ and $t_k = t_0 + k\pi/\Omega$ and $t_0$ is known.

These $x\ind{j}(t)$'s are mixed before being input to $I$ TEMs. The mixing is described by a matrix $\mat{A} \in \mathbb{R}^{I\times J}$:
\begin{equation}
\label{eq: y_j linear combination of x_i's}
    y\ind{i}(t) = \sum_{j = 1}^N a_{ij} x\ind{j}(t).
\end{equation}
Here, $a_{ij}$ is the element in the $i^{th}$ row and $j^{th}$ column of $\mat{A}$ and $y\ind{i}(t)$ denotes the $i^{th}$ output of the mixing.

Each of these signals $y\ind{i}(t)$ is then sampled using a time encoding machine TEM$\ind{i}$. Every TEM$\ind{i}$ acts as an IF-TEM, as defined above, and produces spikes at times $\spikes{i}$. These spike times will form the sample set output by the machines. The reconstruction algorithm will make use of the fact that the spike times place constraints on the signals $y\ind{i}(t)$ and thus indirectly on the signals $x\ind{j}(t)$.

\subsection{Notation and Constraints}
\label{sec: Setup constraints}
We denote our collection of continuous signals $x\ind{j}(t)$, $j = 1\cdots J$, as a ``vector signal" $\vect{x}(t)$, where $x\ind{j}(t)$ is the $j$\textsuperscript{th} element of $\vect{x}(t)$. %Thus, sampling $\vect{x}(t)$ at time $t_0$ gives a vector $\vect{x}(t_0)\in\mathbb{R}^N$.
Similarly, we denote $\vect{y}(t)$ to be the collection of signals $y\ind{i}(t)$, and thus we rewrite~\eqref{eq: y_j linear combination of x_i's} as $\vect{y}(t) = \mat{A}\vect{x}(t)$.

Our setup assumes that there exists $\Omega$ and $K$, such that~\eqref{eq: condition on x's} is satisfied.
It directly follows that the signals $y\ind{i}(t)$, which are linear combinations of the $x\ind{j}(t)$'s also satisfy~\eqref{eq: condition on x's}.

% Each of the $y\ind{i}(t)$'s therefore has $K$ degrees of freedom. Therefore, for each of them to be perfectly recovered, we expect need $K$ samples from each.

The matrix $\mat{A}$ is assumed to be a known $I\times J$ matrix such that every $J$ rows of $\mat{A}$ are linearly independent. This ensures, among other things, that $\mat{A}$ is rank $J$ and therefore has a pseudo-inverse, and that the setup cannot be seperated into several independendent subnetworks.

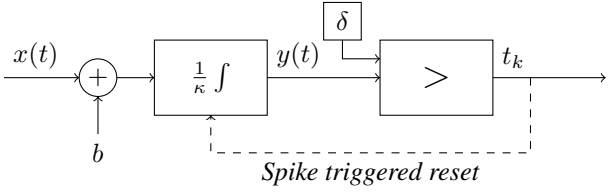
\begin{figure}[tb]
	\begin{minipage}[b]{0.85\linewidth}
		\centering
        \begin{tikzpicture}[scale = 1]

% draw input line
    \tikzmath{
        \inputstartx = 0;
        \inputstarty = 0.4;
        \inputlen = 1;
    }
    \draw[->] (\inputstartx, \inputstarty) node[anchor = south west] {$x(t)$} -- (\inputlen, \inputstarty);

% draw plus
    \tikzmath{
        \circleradius = 0.25;
        \circlecenterx = \inputstartx+\inputlen+\circleradius;
        \circlecentery = \inputstarty;
    }
    \draw (\circlecenterx, \circlecentery) circle (\circleradius) node {$+$};

% draw b to plus
    \tikzmath{
        \biaslinelen = 0.5;
        \biaslinex = \circlecenterx;
        \biasliney = \circlecentery -  \circleradius - \biaslinelen;
    }
    \draw[->] (\biaslinex, \biasliney) node[below] {$b$} --  (\biaslinex, \biasliney+ \biaslinelen);

% draw link to integrator box
    \tikzmath{
        \tointlinex = \circlecenterx + \circleradius;
        \tointliney = \circlecentery;
        \tointlinelen = 0.5;
    }
    \draw[->] (\tointlinex, \tointliney) -- (\tointlinex + \tointlinelen, \tointliney);

% draw integrator box
    \tikzmath{
        \integboxheight = 1;
        \integboxwidth = 1.5;
        \integboxx = \tointlinex + \tointlinelen;
        \integboxy = \tointliney - \integboxheight/2;
    }
    
    \draw (\integboxx, \integboxy) rectangle (\integboxx+\integboxwidth, \integboxy+\integboxheight) node[midway] {$\frac{1}{\kappa}\int$};

% draw link to comparator box
    \tikzmath{
        \tocomplinex = \integboxx + \integboxwidth;
        \tocompliney = \tointliney;
        \tocomplinelen = 1.5;
    }
    \draw[->] (\tocomplinex, \tocompliney) node[anchor = south west] {$y(t)$} -- (\tocomplinex + \tocomplinelen, \tocompliney);
    
    % draw comparator box
    \tikzmath{
        \compboxheight = 1;
        \compboxwidth = 1.5;
        \compboxx = \tocomplinex + \tocomplinelen;
        \compboxy = \tocompliney - \compboxheight/2;
    }
    
    \draw (\compboxx, \compboxy) rectangle (\compboxx+\compboxwidth, \compboxy+\compboxheight) node[midway] {\Large $>$};
    
    % draw delta link to comparator box
    \tikzmath{
        \deltalinkx = \compboxx;
        \deltalinky = \compboxy + 0.75\compboxheight;
        \deltalinkwidth = 0.5;
        \deltalinkheight = 0.25;
    }
    \draw[->] (\deltalinkx - \deltalinkwidth, \deltalinky + \deltalinkheight) -- (\deltalinkx- \deltalinkwidth, \deltalinky) -- (\deltalinkx, \deltalinky);
    
    % draw delta box
    \tikzmath{
        \deltaboxwidth = 0.5;
        \deltaboxheight = 0.5;
        \deltaboxx = \deltalinkx - \deltalinkwidth - \deltaboxwidth/2;
        \deltaboxy = \deltalinky + \deltalinkheight ;
    }
    
    \draw (\deltaboxx, \deltaboxy) rectangle  node {$\delta$} (\deltaboxx + \deltaboxwidth, \deltaboxy+ \deltaboxheight);
    
    % draw comp link to output
    \tikzmath{
        \tooutlinex = \compboxx + \compboxwidth;
        \tooutliney = \tocompliney;
        \tooutlinelen = 1.5;
    }
    \draw[->] (\tooutlinex, \tooutliney) node[anchor = south west] {$t_k$} -- (\tooutlinex + \tooutlinelen, \tooutliney);
    
    % draw reset line
    \tikzmath{
        \feedbacklinex = \tooutlinex +0.5;
        \feedbackliney = \tooutliney;
        \feedbacklineheight = 1;
        \feedbacklineendy = \integboxy;
        \feedbacklineendx = \integboxx + \integboxwidth/2;
    }
    \draw[dashed, ->] (\feedbacklinex, \feedbackliney) -- (\feedbacklinex , \feedbackliney- \feedbacklineheight) -- node[below] {\textit{Spike triggered reset}} (\feedbacklineendx , \feedbackliney - \feedbacklineheight) --    (\feedbacklineendx, \feedbacklineendy);
\end{tikzpicture}
	\end{minipage}
	\vspace{-2em}
	\caption{Circuit of a Time Encoding Machine.}
	\label{fig:TEM circuit}
	\end{figure}

As for the TEMs used to sample the $y\ind{i}(t)$'s, they behave like IF-TEMs as depicted in Fig.~\ref{fig:TEM circuit} and defined above.
We assume that the TEM$\ind{i}$'s have different but known parameters $\kappa\ind{i}$, $\delta\ind{i}$ and $b\ind{i}$, and that the spike times are known. 
%In other words, we have individual spike streams $\spikes{i}$ from each TEM$\ind{i}$. 
These spike streams satisfy
% \begin{equation}
   $\int_{t_\ell\ind{i}}^{t_{\ell+1}\ind{i}} \left( y\ind{i}(u)+b\ind{i}\right) \, du = 2\kappa\ind{i} \delta\ind{i}.$
% \end{equation}

% We also define the spiking rate of a TEM as follows.
% \begin{definition}
% The \emph{spiking rate} $r\ind{j}$ of TEM$\ind{j}$ is defined to be the average number of spikes emitted per unit time.
% \begin{equation}
%     r\ind{j} = \lim_{t\rightarrow \infty} \frac{n\ind{j}\left(\left[-t,t\right]\right)}{2t},
% \end{equation}
% where $n\ind{j}\left(\left[-t,t\right]\right)$ denotes the number of spikes that are generated by TEM$\ind{j}$ over the interval $\left[-t,t\right]$.\\
% If we let $\bar{y}\ind{j}$ be the mean value of the input $y\ind{j}(t)$, it can be shown that $r\ind{j} = \max\left(0,\pi\left(b\ind{j}+\bar{y}\ind{j}\right)/\left(2\kappa\ind{j}\delta\ind{j}\right)\right)$.
% \end{definition}

\subsection{Conditions for reconstructibility}
In previous work, the constraints for reconstructability were written in terms of the bandwidth $\Omega$, the signal bound $c$, and the parameters of the TEMs $\kappa, \delta$ and $b$~\cite{lazar2004perfect,adam2019sampling,lazar2005multichannel,lazar2010population,lazar2008faithful}. These constraints arise because of a relationship between the spiking rate of each machine $r\ind{i}$ and the parameters of the machine. In fact, if $b>c$, we get
% \begin{equation}
    $r\ind{i} \geq \pi\left(b-c\right)/\left(2\kappa\delta\right)$.
% \end{equation}
Therefore, placing a constraint on the parameters of the machines effectively places a constraint on the sampling rate of the machines.
In this paper, we will directly place constraints on the spiking rate of the machines, for two reasons.
 \begin{enumerate}
     \item Previous work required that the bias $b$ be such that $b>c\geq |x(t)|, \forall t$. This requires knowing the maximal values that the input signal will obtain and setting $b$ accordingly. This is difficult to do in practice, and setting a large $b$ induces a high spiking rate on the machines. Therefore, we prefer to make no assumption on $b$, %itself, even allowing it to be zero, 
     and rather constrain the spiking rate of the machines.
    \item Bounds that depend on $\kappa$, $\delta$, $b$ and $c$ are not tight and placing constraints that depend on the spiking rates $r\ind{i}$ provides a tighter bound on the bandwidth.
 \end{enumerate}

We can now state our main result.
\begin{theorem}
\label{thm: N signal M channel reconstructibility}
Assume $\vect{x}(t)$ is a vector of signals $x\ind{j}(t), j = 1\cdots J$ satisfying~\eqref{eq: condition on x's} where the $c_{jk}$'s are drawn from a Lipschitz continuous probability distribution. Now let $\mat{A} \in \mathbb{R}^{I \times J}$ have every $J$ rows linearly independent, and $\vect{y}(t) = \mat{A}\vect{x}(t)$. Then let each $y\ind{i}(t)$ be sampled using an IF-TEM which starts sampling at $t_0^{(i)}$ with a known initial condition $\zeta_0^{(i)} = -\kappa^{(i)}\delta^{(i)}$ and emits  spike times $\spikes{i}$.
The input $\vect{x}(t)$ is exactly determined by the spike times if
\begin{equation}
    \label{eq: spiking rate condition for reconstructibility}
    \sum_{i=1}^I \min \left(n\ind{i}_{\mathrm{spikes}}, K \right) > JK.
\end{equation}

\end{theorem}
% \begin{proof}
% A manuscript with the proof will be available online and cited in the camera-ready version of this paper.
% \end{proof}

To grasp the intuition behind the condition in~\eqref{eq: spiking rate condition for reconstructibility}, let us first consider the following relaxed condition:
\begin{equation}
    \label{eq: spiking rate condition - relaxes}
    \sum_{i=1}^I n\ind{i}_{\mathrm{spikes}} > JK.
\end{equation}

This condition is necessary if~\eqref{eq: spiking rate condition for reconstructibility} holds but it is not sufficient for~\eqref{eq: spiking rate condition for reconstructibility} to hold. It requires that the total number of spikes of all machines is greater than $JK$. 
Such a condition seems intuitive: $K$ denotes the number of degrees of freedom of each $x\ind{j}(t)$. Therefore, to reconstruct $\vect{x}(t)$, one needs to recover $JK$ degrees of freedom and thus needs at least as many spike times in total.

Considering again the initial condition in~\eqref{eq: spiking rate condition for reconstructibility}, the $\min$ term highlights the fact that $K$ is the highest ``useful'' number of samples when one performs noiseless sampling. In fact, assume TEM$\ind{i}$ emits $n\ind{i}_{\mathrm{spikes}}$ for an input satisfying~\eqref{eq: condition on x's}. If $n\ind{i}_{\mathrm{spikes}}>K$, the information encoded is no greater than the information encoded when the spiking rate is $n\ind{i}_{\mathrm{spikes}} = K$.

The condition in~\eqref{eq: spiking rate condition for reconstructibility} implies that
the spikes $t_\ell^{(i)}$ of TEM$\ind{i}$ do not need to be able to reconstruct the input $y\ind{i}(t)$ for reconstructibility of the $x\ind{j}(t)$'s to be guaranteed.
It also implies that machines that spikes too little can be compensated for by other machines that spike more often, but only up to a certain limit, as is shown by the $\min (n\ind{i}_{\mathrm{spikes}}, K)$ term.

A proof of Theorem~\ref{thm: N signal M channel reconstructibility} is based on matrix recovery using bilinear measurements and is detailed in~\cite{pacholska2020matrix}.

\section{Reconstruction Algorithm}
As we did previously~\cite{adam2019sampling}, we use a projection onto convex sets algorithm to reconstruct a signal from its spike times.

\begin{definition}
	The \textit{projection onto convex sets (POCS) method} obtains a solution for $x$, called $\hat{x}$, by alternately projecting on each of the convex sets $\mathcal{C}_1, \mathcal{C}_2,\cdots,\mathcal{C}_N$, using operators $\mathcal{P}_1,\mathcal{P}_2,\cdots,\mathcal{P}_N$. Here, we assume that $\exists N \in \mathbb{N}$ such that the element $x$ we are looking for lies in the intersection of $N$ known convex sets $\mathcal{C}_1, \mathcal{C}_2,\cdots,\mathcal{C}_N$ which are subsets of a Hilbert space $X$. 
	\end{definition}

The POCS algorithm is known to converge to a fixed point which lies in the intersection of the sets at hand $\bigcap_{n=1}^N\mathcal{C}_n $~\cite{bauschke1996projection, thao2019pseudo}. Thus, if the intersection of the sets consists of a single element, then the algorithm converges to the \textit{correct} solution.
	
Theorem~\ref{thm: N signal M channel reconstructibility} stated that, if~\eqref{eq: spiking rate condition for reconstructibility} is satisfied, the solution $\vect{x}(t)$, and thus $\vect{y}(t) = \mat{A}\vect{x}(t)$, is unique. We will therefore set up a POCS algorithm to first recover $\vect{y}(t)$ and then $\vect{x}(t)$.
	
To recover $\vect{y}(t)$, we define three convex sets: the set $\setom$ of collections of $I$ functions formed using a sum of $L$ sincs as in~\eqref{eq: condition on x's}, the set $\set{spikes}$ of collections of $I$ functions that satisfy the constraints that are set by the spike times of each machine $\spikes{i}$ and the set $\set{A}$ of collections of functions $\vect{y}(t)$ which can be written $\vect{y}(t) = \mat{A}\vect{x}(t)$.
	
\begin{lemma}
	The intersection $\setom\cap\set{spikes}\cap\set{A}$ is the set of solutions $\vect{y}(t)$, given spike times $t\ind{i}_k$ and mixing matrix $\mat{A}$.
\end{lemma}
\begin{proof}
	It is easy to see that a solution $\vect{\hat{y}(t)}$ lies in $\setom\cap\set{spikes}\cap\set{A}$. Now assume that $\vect{\hat{y}}(t) \in \setom\cap\set{spikes}\cap\set{A}$. Then $\exists\, \vect{\hat{x}}(t)$ $2\Omega$-bandlimited such that $\vect{\hat{y}}(t) = \mat{A}\vect{\hat{x}}(t)$ and $\vect{\hat{y}}(t)$ produces the obtained spike times. Therefore $\vect{\hat{y}}(t)$ is a solution to the input of the machines.
\end{proof}
	
Each of these sets is convex, therefore, we define operators $\projom$, $\proj{spikes}$, and $\proj{A}$ that project orthogonally onto $\setom$, $\set{spikes}$ and $\set{A}$, respectively. Then, we alternately apply these projection operators to an initial estimate, and, since the intersection is unique, we converge to the correct solution.
	
To project onto $\setom$, we convolve the input with a sinc of bandwidth $\Omega$, sample the obtained signal at values $t_k$ and use the values as amplitudes of the sincs located at $t_k$:
\begin{equation}
    \projom\ind{i} \left( \vect{\hat{y}}(t)\right) = \sum_{k=1}^K \hat{y}_{\Omega}\ind{i}(t_k) \mathrm{sinc}_\Omega(t-t_k),
\end{equation}
where  $\hat{y}_{\Omega}\ind{i}(t) = \hat{y}\ind{i}(t)\, *\, \mathrm{sinc}_\Omega (t)$.
To project onto $\set{spikes}$, we define $\proj{spikes}$ to act on each row of $\vect{\hat{y}}(t)$ individually:
\begin{equation}
    \proj{spikes}\ind{i} \left( \vect{\hat{y}}(t)\right) = \hat{y}\ind{i}(t) + \sum_{\ell=1}^L q_\ell\ind{i} \frac{\mathbbm{1}_{[t^{(i)}_\ell, t^{(i)}_{\ell+1})}(t)}{t^{(i)}_{\ell+1}-t^{(i)}_\ell}
\end{equation}
where $q_\ell\ind{i} = \int_{t_\ell\ind{i}}^{t_{\ell+1}\ind{i}}\left( \hat{y}\ind{i}(u) - y\ind{i}(u)\right) \, du$, and $\mathbbm{1}_{[a,b)}(t)$ is the indicator function over $[a,b)$. 

In words, for each $\hat{y}\ind{i}(t)$, $\proj{spikes}$ adds rectangles over the intervals $\left[t_\ell\ind{i}, t_{\ell+1}\ind{i} \right]$ with appropriate weights to satisfy the constraints set by $\spikes{i}$.
    	
Now, to project $\vect{\hat{y}}(t)$ onto the set $\set{A}$, we let
\begin{equation}
    \proj{A} \left(\vect{\hat{y}}(t)\right) = \mat{A}\left(\mat{A}^T\mat{A} \right)^{-1}\mat{A}^T\vect{\hat{y}}(t).
\end{equation}

Thus, our reconstruction algorithm runs iteratively and computes new values $\vect{y_\ell}(t)$ of the originally sampled signals:
\begin{align}
    \vect{y}_0(t) &= 0,\\
    \vect{y}_{m+1}(t) &= \projom \left( \proj{A} \left(\proj{spikes}\left(\vect{y}_m(t)\right) \right)\right).
\end{align}
    
In the end, we set $\vect{x_m}(t) = \left(\mat{A}^T\mat{A} \right)^{-1}\mat{A}^T \vect{y_m}(t)$.
    
Given that this is a POCS algorithm and that our theorem states uniqueness, $\vect{y_m}(t)$, and therefore $\vect{x_m}(t)$, will converge to the correct solution as $m\rightarrow \infty$.

\section{Simulations}
\begin{figure}[tb]
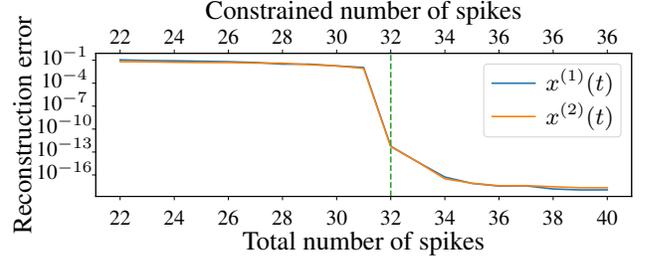

	\centering
    \def\svgwidth{0.95\columnwidth}
    % \subfile{Figures/BetterPerformance.tex}
    \subfile{Figures/Figure3.tex}
    \vspace{2pt}
	\caption{Reconstruction error of two signals $x\ind{1}(t)$ and $x\ind{2}(t)$ when sampled using three TEMs, as the number of spikes of two machines are fixed at 12 and 8, and the spiking rate of the third machine varies by varying the bias $b$. The dashed green line marks the needed spiking rate for reconstructibility.}
	\label{fig: Performance improves}
\end{figure}

In Fig.~\ref{fig: Performance improves}, we want to study the convergence of our algorithm with respect to~\eqref{eq: spiking rate condition for reconstructibility} being satisfied. To do this, we fix a system with 2 input signals $x\ind{1}(t)$ and $x\ind{2}(t)$ and 3 IF-TEMs. Both inputs satisfy~\eqref{eq: condition on x's} with $\Omega = \pi$ and $L=16$ so that the critical number of samples is $16\times 2=32$ (shown in green in Fig.~\ref{fig: Performance improves}). We define a mixing matrix $\mat{A}$ that has every 2 rows linearly independent. We then randomly generate 100 signals $\vect{x}(t)$ and sample them using the IF-TEMs\footnote{The code that performs multi-channel encoding and decoding is available at \href{https://github.com/karenadam/Multi-Channel-Time-Encoding}{https://github.com/karenadam/Multi-Channel-Time-Encoding}.} as described in~\cite{adam2019multi-channel}. We  study the mean-squared error with respect to the number of spikes emitted, when using a closed form reconstruction algorithm similar to the one devised in~\cite{adam2019multi-channel}: two machines have fixed number of spikes of 12 and 8 and the third has a number of spikes which is varied by varying the bias of the machine~\cite{karen_adam_2020_3666801}. Notice how the reconstruction error decreases sharply once the critical number of spikes (in green) is reached.

\section{Conclusion}
We have proposed a setup of multi-signal multi-channel time encoding using integrate-and-fire neurons and a known mixing matrix. In our scenario, the bandlimited input signals can be reconstructed if the  overall spiking rate of the machines is higher than the Nyquist rate. 
We then provided an iterative reconstruction algorithm and included simulation results to show that the algorithm converges to the correct solution under the proper constraints.

In this paper, we assumed that the input signals are all bandlimited  with the same bandwidth. The setup could easily be extended to having multiple bandpass signals with different frequency supports, and the conditions for reconstruction would remain similar. We also hope to extend the setup to understand encoding and decoding of non-bandlimited signals.

\section{Acknowledgements}
The authors would like to thank Thao Nguyen for his input on the convergence of POCS algorithms and Michalina Pacholska for discussions about uniqueness of the reconstruction.

\bibliographystyle{IEEEbib}

\bibliography{main}
% \printbibliography

\end{document}